\documentclass{article}
\usepackage[T2A]{fontenc}
\usepackage[cp866]{inputenc} 
\usepackage[english]{babel}
\usepackage[tbtags]{amsmath}
\usepackage{amsfonts,amssymb,mathrsfs,amscd,comment}
\usepackage{umnbib} 

\newtheorem{theorem}{Theorem}[section]

\newtheorem{definition}[theorem]{Definition}

\numberwithin{equation}{section}
\DeclareMathOperator{\id}{id}

\begin{document}

\title{\bf\large\MakeUppercase{%
Scalene Yang--Baxter maps and Lax triples
}}

\author{S. Konstantinou-Rizos and T. Kouloukas}
\date{}

\maketitle

\makeatletter
\renewcommand{\@makefnmark}{}
\makeatother
\footnotetext{The work of SKR was supported by the Theoretical Physics and Mathematics Advancement Foundation ``BASIS'' (Grant No. 25-7-2-10-1).}

 In this work, we introduce a generalisation of the set-theoretical Yang--Baxter equation \cite{Buchstaber, Drinfeld, Veselov}, and study the relation of its solutions to matrix refactorisation problems. The proposed term `scalene' refers to the asymmetric action of the three maps that appear in the equation. Such maps have already appeared in the literature \cite{Adler-Enc, Kassotakis-Proc}.

\begin{definition} Let $(S,R,T)$ be maps $S:\mathcal{X}_1\times\mathcal{X}_2\rightarrow \mathcal{X}_1\times\mathcal{X}_2$, $R:\mathcal{X}_1\times\mathcal{X}_3\rightarrow \mathcal{X}_1\times\mathcal{X}_3$ and $T:\mathcal{X}_2\times\mathcal{X}_3\rightarrow \mathcal{X}_2\times\mathcal{X}_3$. The maps $(S,R,T)$ are called \textit{scalene Yang--Baxter maps}, if they satisfy the \textit{scalene Yang--Baxter equation}:
\begin{equation}\label{scalene-YB}
    S^{12}\circ R^{13}\circ T^{23}=T^{23}\circ R^{13}\circ S^{12},
\end{equation}
where $S^{12},R^{13},T^{23}:\mathcal{X}_1\times\mathcal{X}_2\times\mathcal{X}_3\rightarrow \mathcal{X}_1\times\mathcal{X}_2\times\mathcal{X}_3$ denote the maps 
$S\times\id_{\mathcal{X}_3}$, $\id_{\mathcal{X}_1}\times T$, $\pi^{12}\circ(\id_{\mathcal{X}_2}\times R)\circ\pi^{12}$ respectively,  
$\id_{\mathcal{X}_i}$ is the identity map on $\mathcal{X}_i$, $i=1,2,3$, and $\pi^{12}$ the permutation map $\pi^{12}(x_1,x_2,x_3)=(x_2,x_1,x_3).$
\end{definition}
Equation \eqref{scalene-YB} is a generalisation of the Yang--Baxter equation: for $\mathcal{X}_1=\mathcal{X}_2=\mathcal{X}_3$, we obtain the entwining Yang--Baxter equation \cite{Kouloukas-2}. If additionally $S\equiv R\equiv T$, then we obtain the set-theoretical Yang--Baxter equation.

{\em Parametric scalene Yang--Baxter maps} are scalene Yang--Baxter maps $(S,R,T)$ acting on an additional invariant parameter space, i.e.  $\mathcal{X}_i=\mathbb{X}_i\times \mathbb{I}_i$, where  $\mathbb{X}_i$,  $\mathbb{I}_i$, $i=1,2,3,$ denote the sets of variables and parameters respectively, which preserve the parameters. That is, 
$S((x,a),(y,b))\mapsto ((u_S,a),(v_S,b))$, $R((x,a),(y,b))\mapsto ((u_R,a),(v_R,b))$, $T((x,a),(y,b))\mapsto ((u_T,a),(v_T,b))$. Following \cite{Veselov}, we keep the parameters separately and denote such maps as $S_{a,b}:\mathbb{X}_1 \times \mathbb{X}_2 \rightarrow \mathbb{X}_1 \times \mathbb{X}_2,$ with $S_{a,b}:(x,y) \mapsto (u_S,v_S)$, and similarly for  $R_{a,b}$ and $T_{a,b}$.


\begin{definition}\label{lax-def}
    Let $\mathbb{K}$ be a field of zero characteristic and ${\rm L}_i: \mathcal{X}_i \times \mathbb{K} \to \mathrm{Mat}_{n \times n}(\mathbb{K})$, $i=1,2,3$. The triad $({\rm L}_1,{\rm L}_2,{\rm L}_3)$ is called a Lax triple for maps $(S,R,T)$, with $S(x,y)=(u_S,v_S)$, $R(x,y)=(u_R,v_R)$ and $T(x,y)=(u_T,v_T)$, if the latter satisfy the matrix refactorisation equations 
\begin{equation}\label{Lax-triad}
    {\rm L}_1(u_S){\rm L}_2(v_S)={\rm L}_2(y){\rm L}_1(x),~
       {\rm L}_1(u_R){\rm L}_3(v_R)={\rm L}_3(y){\rm L}_1(x),~
       {\rm L}_2(u_T){\rm L}_3(v_T)={\rm L}_3(y){\rm L}_2(x),
\end{equation}for all $\lambda\in\mathbb{K}$. Here, in ${\rm L}_i$ we omitted the dependence on the spectral parameter, i.e. ${\rm L}_i(x,\lambda)\equiv{\rm L}_i(x)$.
\end{definition}


\begin{theorem}\label{trifactorisation}
    If maps $(S,R,T)$ satisfy system \eqref{Lax-triad} and the matrix trifactorization equation  ${\rm L}_1(\hat{x}_1){\rm L}_2(\hat{x}_2){\rm L}_3(\hat{x}_3)={\rm L}_1(\tilde{x}_1){\rm L}_2(\tilde{x}_2){\rm L}_3(\tilde{x}_3)$ implies only the trivial solution $\hat{x}_i=\tilde{x}_i$, $i=1,2,3$, for any $\lambda\in\mathbb{K}$, then  $(S,R,T)$ are scalene Yang--Baxter maps.
\end{theorem}

Definition \ref{lax-def} and Theorem \ref{trifactorisation} extend to the parametric case when  ${\rm L}_i(x)={\rm L}_i(x', a)$, for $x' \in \mathbb{X}_i$, $a \in \mathbb{I}_i$.
\vspace{0.2cm}

\noindent\textbf{Scalene Yang--Baxter maps of $H$-type.} Let $\mathbb{X}_1=\mathbb{X}_2=\mathbb{C}^2$, $\mathbb{X}_3=\mathbb{C}$ and $\mathbb{I}_i=\mathbb{K}=\mathbb{C}$. Consider  the matrices {\small\begin{equation}\label{Lax-scal-KdV}
        {\rm L}_1(x_1,x_2,a)=\begin{pmatrix}x_1 & x_1x_2+a-\lambda\\ 1 & x_2\end{pmatrix},\quad
        \quad 
        {\rm L}_2(x,a)={\rm L}_3(x,a)=\begin{pmatrix}x & x^2+a-\lambda\\ 1 & x\end{pmatrix}.
    \end{equation}}
The matrices ${\rm L}_1$ and ${\rm L}_2$ are Lax matrices for a KdV lift \cite{Kouloukas} and the Adler map \cite{Adler} (which is $H_V$ map in \cite{PSTV}), respectively. Now, the system \eqref{Lax-triad} is equivalent to the maps 
    {\footnotesize \begin{equation}\label{scal-KdV}
        S_{a,b}((x_1,x_2),y)=\left(\left(y-\frac{a-b}{x_1+y},x_1-x_2-\frac{a-b}{x_1+y}\right),x_2+\frac{a-b}{x_1+y}\right),   T_{a,b}(x,y)=\left(y-\frac{a-b}{x+y},x+\frac{a-b}{x+y}\right).
    \end{equation}}Moreover, the matrix trifactorisation problem ${\rm L}_1(u_1,u_2,a){\rm L}_2(v,b){\rm L}_2(w,c)= \linebreak{\rm L}_1(x_1,x_2,a){\rm L}_2(y,b){\rm L}_2(z,b)$ implies the unique solution  $u_1=x_1, u_2=x_2,v=y,w=z$.  Therefore, from Theorem \eqref{trifactorisation}  we conclude that the maps $(S_{a,b},S_{a,b},T_{a,b})$ from \eqref{scal-KdV} are parametric scalene Yang--Baxter maps. 

    Next, we consider the following matrices:
    {\small\begin{equation}\label{Lax-scal-relat}
        {\rm L}_1(x_1,x_2,a)=\begin{pmatrix}\lambda & ax_1\\ \frac{a}{x_2} & \lambda\end{pmatrix},\quad
          {\rm L}_2(x,a)= {\rm L}_3(x,a)=\begin{pmatrix}\lambda & ax\\ \frac{a}{x} & \lambda\end{pmatrix}.
    \end{equation}}
   System \eqref{Lax-triad} is equivalent to 
   {\footnotesize \begin{equation}\label{scal-relat}
        S_{a,b}((x_1,x_2),~y)=\left(\left(\frac{ax_1+by}{bx_2+ay}y, ~\frac{ax_1+by}{bx_2+ay}y\cdot\frac{ x_2}{x_1}\right),~\frac{ax_1+by}{bx_2+ay}x_2\right),~ T_{a,b}(x,y)=\left(\frac{ax+by}{bx+ay}y,~\frac{ax+by}{bx+ay}x\right).
    \end{equation}}
    The map $T_{a,b}$ represents  elastic relativistic collisions \cite{Kouloukas-2017}, and is related to the modified  KdV lattice equation ($H_3^A$ map in \cite{PSTV}). Similarly to the previous case, Theorem \eqref{trifactorisation} implies that  $(S_{a,b},S_{a,b},T_{a,b})$, defined by \eqref{scal-relat}, are parametric scalene Yang--Baxter maps. 
\vspace{0.2cm}

\noindent \textbf{NLS type scalene maps.} Let $\mathcal{X}_1=\mathbb{C}^3$, $\mathcal{X}_2=\mathbb{C}^2$, $\mathcal{X}_3=\mathbb{C}$ and $\mathbb{K}=\mathbb{C}$. Consider the matrices
   {\small \begin{equation}\label{Lax-scal-NLS}
        {\rm L}_1(x_1,x_2,x_3)=\begin{pmatrix}x_1x_2+x_3+\lambda & x_1\\ x_2 & 1\end{pmatrix},~
        {\rm L}_2(x_1,x_2)=\begin{pmatrix}x_2+\lambda & x_1\\ \frac{1}{x_1} & 0\end{pmatrix},~ 
        {\rm L}_3(x)=\begin{pmatrix}x & 0\\ 0 & \frac{1}{x}\end{pmatrix}.
    \end{equation}}
     The matrices ${\rm L}_1$ and  ${\rm L}_2$ are certain Darboux matrices for the AKNS (or NLS) system \cite{SPS}. After substitution of \eqref{Lax-scal-NLS} into \eqref{Lax-triad}, the latter is equivalent to the maps\vspace{-0.2cm}
   {\small \begin{subequations}\label{scalene-NLS}
        \begin{align}
        &S((x_1,x_2,x_3),~(y_1,y_2))=\left(\left(y_1(y_2-x_3)+\frac{y_1^2}{x_1},~\frac{1}{y_1},~x_3\right),~\left(x_1,~x_1x_2+x_3-\frac{y_1}{x_1}\right)\right),\\
        &R((x_1,x_2,x_3),~y)=\left(\left(x_1y^2,~\frac{x_2}{y^2},~x_3\right),~y\right),\quad T((x_1,x_2),y)=\left((x_1y^2,x_2),~y\right). 
    \end{align}
    \end{subequations}}
     Moreover, equation ${\rm L}_1(u_1,u_2,u_3){\rm L}_2(v_1,v_2){\rm L}_3(w)= {\rm L}_1(x_1,x_2,x_3){\rm L}_2(y_1,y_2){\rm L}_3(z)$ implies \linebreak$u_i=x_i,v_i=y_1,w_i=z_i$, $i\in\{1,2,3\}$. Therefore, according to Theorem \eqref{trifactorisation}, the maps $(S,R,T)$ are scalene Yang--Baxter maps. None of the maps $(S,R,T)$ individually is a Yang--Baxter map, but the three of them together satisfy the scalene Yang--Baxter equation.

All the maps presented in this paper generate integrable transfer dynamics \cite{Veselov}. Further-more, they correspond to integrable lattice equations and admit soliton solutions, which will be presented in a future work.

\vspace{.5cm}

\noindent{\small
\textbf{S. Konstantinou-Rizos}\\
P.G. Demidov Yaroslavl State University, Yaroslavl, Russia,
\texttt{skonstantin84@gmail.com}

\noindent
\textbf{T. Kouloukas}\\
London Metropolitan University, London, UK, \texttt{t.kouloukas@londonmet.ac.uk}
}

\end{document}